&&&&&&&&&&&&&&&&&&&&&&&&&&&&&&&&&&&&&&&&&&&&&&&&&&&&&&&&&&&&&&&&&&&&&&&&&&&&&&&&
\documentstyle[12pt]{article}

\begin{document}
\rightline{ULB-TH 14/94}

{}~\\
{}~\\
\begin{center}
{\huge {\bf Self-dual solutions of 2+1 Einstein gravity with a
negative
cosmological constant}\footnote{To appear in the Proceedings of the
 Meeting `` The
Black Hole: 25 years after'' held in Santiago in January 1994.}}\\
\vspace{2.5cm}
{\large Olivier Coussaert$^{a}$  } and
\vspace{1cm}
{\large Marc Henneaux$^{a,b}$ }\\\vspace{1.5cm}
{}~$^a$ {\em Facult\'e des Sciences, Universit\'e
Libre de Bruxelles, Campus Plaine C.P. 231, B-1050
Bruxelles, Belgium.} \\
{}~$^b${\em Centro de Estudios Cient\'{\i}ficos de Santiago,
Casilla 16443, Santiago 9, Chile} \\

\vspace{2cm}
\end{center}
\vfill

\eject

\abstract{ All the causally regular
geometries obtained from (2+1)-anti-de Sitter space by
identifications by isometries
of the form $P \rightarrow (\exp 2 \pi \xi) P$, where $ \xi$ is a self-
dual Killing
vector of $so(2,2)$, are explicitely constructed. Their remarkable
symmetry
properties (Killing vectors, Killing spinors) are listed. These solutions
of Einstein
gravity with negative cosmological constant are also shown to be
invariant under the
string duality transformation applied to the angular translational
symmetry $ \phi
\rightarrow \phi + a $. The analysis is made particularly convenient through
the
 construction of
 {\em global} coordinates adapted to the identifications.}

\section{Introduction.}

It has been discovered recently \cite{BTZ} that 2+1 Einstein gravity
with a negative
cosmological constant, even though devoid of local degrees of
freedom, allows for
black hole solutions with features quite similar to those occuring in
the standard
four-dimensional theory. The geometry of these black hole solutions
have been
investigated in depth in \cite{BHTZ}, and the remarkable
supersymmetry properties of
the extreme black hole with mass equal to angular momentum, as
well as of the
massless black hole ground state, have been studied in \cite{CH}.
Further analysis of the 2+1 black hole may be found in \cite{xxx}.

The talk presented by one of us (M.H.) at the Santiago meeting was
devoted to a
survey of these results. Since these can be found in the existing
literature, we
shall not repeat them here. Rather, we shall analyse a question that
has come across
in the study of the geometry of the (2+1)-black hole \cite{BHTZ},
namely, that of
determining all ``self-dual'' solutions of Einstein gravity with
a negative cosmological
constant. [ In what sense these solutions are self-dual is defined
precisely in the
next section]. Although these solutions do not belong to the black
hole family, they
exhibit quite interesting geometric features that make them worth of
study. Namely,
they have four Killing vectors, and possess in addition two Killing
spinors.
Futhermore, they are invariant under the string duality
transformations applied to the
angular translational symmetry $\phi \rightarrow \phi + a$.

\section{Self-dual metrics: definition.}

As shown in \cite{BHTZ}, the (2+1)-black hole may be obtained from
anti-de Sitter
space by identifications. These identifications are made as follows.
Any Killing
vectors $\xi$ of the anti-de Sitter metric defines a one-parameter
subgroup of
isometries:

\begin{eqnarray}  \label{2.1} P \rightarrow \exp(t\xi) P
\end{eqnarray}

We shall consider the case where the orbits (\ref{2.1}) are
isomorphic with the real
line. The mappings (\ref{2.1}) for which $t$ is an integer multiple of
a ``basic''
step, taken conventinally as $2 \pi$,

\begin{eqnarray} \label{2.2} P \rightarrow \exp(t \xi) P, \ \ t=0,
\pm 2\pi, \pm
4\pi,... \end{eqnarray}
define a discrete subgroup isomorphic to $Z$.

Since the transformations (\ref{2.2}) are isometries, the quotient
space obtained by
identifying points that belong to a given orbit, inherits from anti-de
Sitter space a
well-defined metric which has the same constant negative curvature
and which is thus
also a solution of the Einstein equations. The identification process
leads therefore
to new solutions of the Einstein equations which differ from anti-de
Sitter space in
their global properties. These solutions are completely characterized
by the group
(\ref{2.2}), which we shall call the identification group. The
identification group,
in turn, is completely characterized by the Killing vector $\xi$ that
generates it.

The Killing vector $\xi$ belongs to the Lie algebra $so(2,2)$. Two
Killing vectors
$\xi$ and $\xi '$ in the same $SO(2,2)$ conjugacy class yield
isomorphic quotient
spaces. Thus, the quotient spaces may be classified by the conjugacy
classes of
$SO(2,2)$ in $so(2,2)$.

The Lie algebra $so(2,2)$ is the direct sum of two copies of
$sl(2,\Re)$,

\begin{eqnarray} \label{2.3} so(2,2)=sl(2,\Re)\oplus sl(2,\Re)
\end{eqnarray}
Let $J_{ab}$ be the Killing vectors of anti-de Sitter metric. In terms
of the
embedding \begin{eqnarray} \label{2.4} -u^2-v^2+x^2+y^2=-l^2
\end{eqnarray} of
anti-de Sitter space in a four-dimensional flat
space of signature \begin{eqnarray} \label{2.5} ds^2=-du^2-
dv^2+dx^2+dy^2,
\end{eqnarray} one has \begin{eqnarray} J_{ab}=x_b
\frac{\partial}{\partial x^a}-x_a
\frac{\partial}{\partial x^b} \end{eqnarray} with $x^a \equiv
(u,v,x,y)$.

The non-vanishing  brackets of the $so(2,2)$-algebra read explicitly:
\begin{equation}
\begin{array}{ll}
{[J_{01},J_{02}]} =-J_{12}  & {[J_{02},J_{03}]}=-J_{23}\\
{[J_{01},J_{03}]} =-J_{13}  & {[J_{02},J_{12}]}=J_{01}\\
{[J_{01},J_{12}]}=J_{02}  & {[J_{02},J_{23}]}=-J_{03}\\
{[J_{01},J_{13}]}=J_{03}  & {[J_{03},J_{13}]}=J_{01}\\
{[J_{03},J_{23}]}=J_{02}  & {[J_{12},J_{13}]}=-J_{23}\\
{[J_{12},J_{23}]}=-J_{13}   & {[J_{13},J_{23}]}=J_{12}
\end{array}
\end{equation}
 One may
take as generators of the first copy $sl(2,\Re)$ (``self-dual''
generators)
\begin{eqnarray} \nonumber \xi_{(0)}=\frac{1}{2}(J_{01}+J_{23})\\
\xi_{(1)}=\frac{1}{2}(J_{02}+J_{13})\\ \nonumber
\xi_{(2)}=\frac{1}{2}(J_{03}-J_{12})
\end{eqnarray} and the generators of the second copy $sl(2,\Re)$
(``anti-self dual''
generators) \begin{eqnarray} \nonumber \eta_{(0)}=\frac{1}{2}
(J_{01}-J_{23})\\
\eta_{(1)}=\frac{1}{2} (J_{02}-J_{13})\\ \nonumber \eta{(2)}=-
\frac{1}{2}
(J_{03}+J_{12}) \end{eqnarray}

The commutation relations between the $\xi$'s and the $\eta$'s
 are: \begin{eqnarray} [\xi_{(0)},\xi_{(1)}]=\xi_{(2)}, \ \
[\xi_{(0)},\xi_{(2)}]=-\xi_{(1)}, \ \ [\xi_{(1)},\xi_{(2)}]=-\xi_{(0)}
\end{eqnarray}
and \begin{eqnarray} [\eta_{(0)},\eta_{(1)}]=\eta_{(2)}, \ \
[\eta_{(0)},\eta_{(2)}]=-\eta_{(1)}, \ \ [\eta_{(1)},\eta_{(2)}]=-
\eta_{(0)}
\end{eqnarray}
Of course, one has also: \begin{eqnarray}
[\xi_{(\mu)},\eta_{(\nu)}]=0. \end{eqnarray}

As shown in \cite{BHTZ}, the Killing vectors used in making the
identifications for
the black hole solutions are neither self-dual nor anti-self dual (for any
value of the
mass and angular momemtum). Nevertheless, identifications by
 self-dual ( or
anti-self-dual) Killing vectors lead to interesting geometries, which
we shall call
self-dual (respectively anti-self-dual).

Since the anti-self-dual case can be obtained from the self-dual one
by the parity
transformation $x^3 \rightarrow -x^3$, which is an isometry, it is
enough to consider
the self-dual case.

There are clearly three subcases to be considered, since there are
only three
inequivalent types of elements in $sl(2,\Re)$: those that are
spacelike (type A),
those are timelike (type B) and those that are lightlike (type C) ( with
the
Killing-Lorentz metric of the $sl(2,\Re)$ algebra). By redefinitions,
one may always
assume: \begin{description} \item{Type A:}  $\xi = a \xi_{(1)}$
\item{Type B:} $\xi =
a \xi_{(0)}$ \item{TypeC:} $\xi=\xi_{(0)}+\xi_{(1)}$ \end{description}

The norms of the Killing vectors ( as tangent vectors to the anti-de
Sitter space,
 with the anti-de Sitter metric) are given by:
\begin{description} \item{Type A:} $\xi . \xi=\frac{a^2}{4} l^2$
\item{Type B:} $\xi
. \xi=-\frac{a^2}{4} l^2$ \item{Type C:} $\xi . \xi = 0$
\end{description} and are
also constant. Thus, the orbits of the Killing vectors are geodesic.
This is quite
remarkable since the anti-de Sitter norm of a generic Killing vector is
in general position
dependent and grows as $r$ as one recedes from the origin.

For comparison with the classification of the $so(2,2)$ elements given
in the
appendix A of \cite{BHTZ}, we observe that type A corresponds to
type $I_b$ with
$\lambda_2=-\lambda_1=a/2$, type B corresponds to type $I_c$
with $b_1=b_2=a/2$ and
type C corresponds to type $II_b$ with $b=0$. [ Note that there is a
misprint in
table $I$ of the appendix of \cite{BHTZ}, where the Killing vector for
type $II_b$
should read $(b-1) J_{01} + (b+1) J_{23} + J_{02} - J_{13}$ instead of
the
incorrect $(b-1) J_{01} + (b-1) J_{23} + J_{02} - J_{13}$.
\footnote{There is another
misprint on page $1521$ of the same paper, where the Casimir
invariants $I_1$ and
$I_2$ are incorrectly expressed in terms of $\omega_{ab}^+$ and
$\omega_{ab}^-$. The
correct expression are $I_1+I_2=\omega_{ab}^+ \omega^{+ab}$ and
$I_1 -I_2 =
\omega^-_{ab} \omega^{-ab}$.}]

\section{(2+1)-anti-de Sitter space as a group manifold.}

One may understand the above properties of the Killing vectors by
recalling that
(2+1)-anti-de Sitter space is the group manifold of $SL(2,\Re)$. This
is most easily
seen by defining new variables \begin{eqnarray}
\alpha=\frac{u+x}{l}, \
\delta=\frac{u-x}{l}, \ \beta=\frac{v-y}{l}, \ \gamma=-\frac{v+y}{l}
\end{eqnarray}
in terms of which the equation $u^2+v^2-x^2-y^2=l^2$ becomes the
unit determinant
condition \begin{eqnarray} \alpha \delta - \beta \gamma =1
\end{eqnarray} for the
matrix
 \begin{eqnarray}
 A=\left( \begin{array}{cc} \alpha & \beta \\ \gamma & \delta
\end{array} \right)
 \end{eqnarray} Furthermore, one verifies straightforwardly that
the anti-de Sitter metric
 coincides with the Killing metric $l^2 {tr(A^{-
1} dA)}^2$ on
the group $SL(2,\Re)$.

Any group manifold $G$ (with the Killing metric) is invariant under
the isometry
group $G_L \times G_R$, where $G_L$ and $G_R$ are respectively the
groups of left and
right translations ( of course, $G_L$, $G_R$ are isomorphic to $G$).
The isometry group is a direct product because left and right
translations commute.
In addition, the infinitisimal generators of $G_L$ (respectively
$G_R$) have constant
norm because, as the Killing metric, they are invariant under right
(respectively
left) translations, which are transitive on $G$. In our case,
$(SL(2,\Re))_L$ is
generated by the self-dual Killing vectors, while $(SL(2,\Re))_R$ is
generated by the
anti-self-dual Killing vectors.

\section{Self dual metrics: construction.}

In order to construct the self-dual metrics of type $A$, we shall
introduce a {\em global}
parametization of anti-de Sitter space adapted to the Killing vector
used in making
the identification. More pecisely, we shall introduce new coordinates
covering the full anti-de Sitter manifold,
in which the
Killing vector $a \xi_{(1)}$ is just $\frac{\partial}{\partial \phi}$.

We shall actually do more, namely, we shall also arrange so that the
anti-self-dual
Killing vector $2 \eta_{(0)}$ is $\frac{\partial}{\partial t}$. [We
choose $2
\eta_{(0)}$ and not any other Killing vector for two reasons: (i)
$\eta_{(0)}$
commutes with $\xi_{(1)}$; (ii) it is everywhere linearly independent
from $\xi_{(1)}$ on
anti-de Sitter space. The factor $2$ is a matter of normalization
convention]. To
simplify the analysis, we start with the case $a=2$. We consider
next the case of an arbitrary $a$.

Along the orbit of $-2 \xi_{(1)}$, one has \begin{eqnarray}
\frac{dv}{d\phi}=y, \
\frac{dy}{d\phi}=v, \ \frac{du}{d\phi}=x, \ \frac{dx}{d\phi}=u
\end{eqnarray} where
$\phi$ is a parameter that distinguishes the points on the same orbit.
Similarly,
along the orbit of $2 \eta_{(0)}$, one has \begin{eqnarray}
\frac{dv}{dt}=u, \
\frac{du}{dt}=-v, \ \frac{dx}{dt}=-y, \ \frac{dy}{dt}=x \end{eqnarray}
We thus introduce the following parametrization of adS
\begin{eqnarray} \label{lum}
u=l(\cosh r \cosh \phi \cos t + \sinh r \sinh \phi \sin t) \nonumber
\\ v=l(\cosh r
\cosh \phi \sin t - \sinh r \sinh \phi \cos t) \nonumber \\ x=l(\cosh
r \sinh \phi
\cos t + \sinh r \cosh \phi \sin t) \nonumber \\ y=l(\cosh r \sinh
\phi \sin t - \sinh
r \cosh \phi \cos t) \end{eqnarray} or equivalently, in terms of the
light-like
variables $\alpha=\frac{u+x}{l}$, $\delta=\frac{u-x}{l}$,
$\beta=\frac{v-y}{l}$,
$\gamma=-\frac{v+y}{l}$, \begin{eqnarray} \label{cgt} \alpha=e^\phi
(\cosh r \cos t +
\sinh r \sin t) \nonumber \\ \beta=e^{-\phi} ( \cosh r \sin t + \sinh r
\cos t)
\nonumber \\ \gamma=-e^{\phi}( \cosh r \sin t - \sinh r \cos t)
\nonumber \\ \delta =
e^{-\phi} ( \cosh r \cos t - \sinh r \sin t) \end{eqnarray}

That this is a good parametrization of anti-de Sitter space can be
seen as follows.
First, any ($\alpha, \beta, \gamma, \delta$) given by (\ref{cgt})
clearly solves the
equation $\alpha \delta - \beta \gamma = 1$. Conversely, let $(
\alpha, \beta,
\gamma, \delta)$ be a solution of $\alpha \delta - \beta \gamma=1$.
Define $\phi$ and
$r$ through \begin{eqnarray} \label{phi} e^{4 \phi} = \frac{\alpha^2
+
\gamma^2}{\beta^2 + \delta^2} \end{eqnarray} and \begin{eqnarray}
\label{rad} \sinh
2r = \alpha \beta + \gamma \delta \end{eqnarray} Because $\alpha
\delta - \beta
\gamma =1$, this last relation implies \begin{eqnarray} \label{**}
{\cosh}^2 2r =
(\alpha^2+\gamma^2) (\beta^2+\delta^2). \end{eqnarray} It remains
to determine $t$.
Since $(\alpha, \gamma)$ belongs to the circle of radius $e^\phi
\cosh^{\frac{1}{2}}
2r$ ($\alpha ^2 + \gamma^2 = e^{2 \phi} \cosh 2r$ by (\ref{phi}) and
(\ref{**})), there
is a unique $t'$ defined by  \begin{eqnarray} \alpha = e^\phi ( \cosh
r \cos t' +
\sinh r \sin t') \nonumber \\ \gamma = - e^\phi ( \cosh r \sin t' -
\sinh r \cos t')
\end{eqnarray} Similarly, there is a unique $t''$ such that
\begin{eqnarray} \beta=
e^{-\phi} ( \cosh r \sin t'' + \sinh r \cos t'') \nonumber \\ \gamma=
e^{-\phi} (
\cosh r \cos t'' - \sinh r \sin t'') \end{eqnarray}

The relations (\ref{rad}) and  $\alpha \delta - \beta \gamma =1$
imply $\sin (t''-t')
+ \cos (t''-t') \sinh 2r = \sinh 2r$ and $\cos (t''-t') -
 \sin (t''-t') \sinh
2r =1$,
i.e. $t''=t'$. We define $t$ by $t=t'=t''$, which shows that given $(
\alpha, \beta,
\gamma, \delta)$ solution of $\alpha \delta - \beta \gamma =1$,
there are unique $(t,
r, \phi)$ in terms of which $ \alpha, \beta, \gamma$ and $\delta$
can be written as
in (\ref{lum}).

In the coordinates ($t, r, \phi$), the anti-de Sitter metric reads
\begin{eqnarray}
\label{3.1} ds^2= -dt^2 + d\phi^2 + 2 \sinh(2r) dt d\phi + d\alpha^2
\end{eqnarray} and the Killing vectors $2 \xi_{(1)}$ and $
2\eta_{(0)}$ are given by
\begin{eqnarray}  2 \xi_{(1)}=\frac{\partial}{\partial \phi} \ , 2
\eta_{(0)}=\frac{\partial}{\partial t} \end{eqnarray} The coordinates
$\phi$ and $r$
range over the entire line.
This is also true for $t$ if one passes to the
universal
covering of adS, as we shall do from now on, \begin{eqnarray}
-\infty < t < +\infty,
\  -\infty < \phi < +\infty, \ -\infty<r<+\infty \end{eqnarray}

If instead of $2 \xi_{(1)}$, one takes as identification vector $a
\xi_{(2)}$, one
needs to make the rescaling $\phi \rightarrow \frac{a}{2} \phi$, in
terms of which
the metric becomes \begin{eqnarray} \label{met_self} ds^2= l^2 (-
dt^2+\frac{a^2}{4}
d\phi^2 + a \sinh 2r dt d\phi + dr^2) \end{eqnarray} and $ a
\xi_{(1)}=\frac{\partial}{\partial \phi}$, $2
\eta_{(0)}=\frac{\partial}{\partial t}$.

The self-dual metric is obtained by making $\phi$ peridodic. Thus, it
is also given
by  (\ref{met_self}), \begin{eqnarray} ds^2=l^2(-dt^2+\frac{a^2}{4}
d\phi^2 - a \sinh
2r dt d\phi + dr^2) \end{eqnarray} but now, $\phi$ is an angle, $-
\infty<t<+\infty, \
0\leq \phi \leq 2 \pi, \  -\infty\leq r\leq +\infty$.

{}From now on, we shall set $l=1$. Note that the self-dual metrics
of type $B$ are obtained by keeping $\phi$ unperiodic and by making $t$
periodic. Similarly,  it is easy to verify
 that the self-dual metric of type
$C$ is given by
\begin{eqnarray}
ds^2=2r^2 dt d\phi + \frac{1}{r^2} dr^2
\end{eqnarray}
For both types $B$ and $C$, there are closed causal curves.
This is not the case for
type $A$, as we now show.

\section{Absence of closed causal curves.}

The self-dual metric (\ref{met_self}) is causaly regular in
 the sense that there is no
closed causal
curves. This is easy to verify because the coordinates $(t,\phi,r)$
provide a global
covering of the manifold. A closed causal curve must fulfill
\begin{eqnarray}
(\frac{dt}{d\lambda})^2 - \frac{a^2}{4} (\frac{d\phi}{d\lambda})^2 -
a \sinh 2r
\frac{dt}{d\lambda} \frac{d\phi}{d\lambda} -
(\frac{dr}{d\lambda})^2 \geq 0
\end{eqnarray} If it closed, $t$ must come back to its original value.
Thus, there
must be a point on the curve where $\frac{dt}{d\lambda}=0$. But
this contradicts the
above inequality ( and the fact that the tangent vector is never zero
since
$\lambda$ is a good parametrization of the curve).

The self-dual space is also geodesically complete (as is anti-de Sitter)
and
singularity-free. Hence, it is a perfectly acceptable solution of the
Einstein
equation. This stationary, axially symmetric solution was missed in
\cite{BTZ,BHTZ}
 because it was assumed from the
very beginning that the rotational Killing vector
$\frac{\partial}{\partial \phi}$
had non constant norm.

\section{Holonomies}

The self-dual metric may also be described in terms of Chern-Simons
holonomies. It is
well known that (2+1)-gravity with a negative cosmological constant
is equivalent to
the Chern-Simons theory with the $SO(2,2)$ gauge group
\cite{Witten}. The self-dual
solution (\ref{met_self}) corresponds to the following $SO(2,2)$ flat
connection,
\begin{eqnarray} A=a [ \cosh r ( J_{02}+J_{13}) - \sinh r (J_{23} +
J_{01}) ] d\phi \nonumber \\ {[-\sinh r (J_{13}-J_{02}) + \cosh r
(J_{01}-J_{23})]} dt \nonumber \\
  -J_{03} dr \end{eqnarray} which, in turn, can be transformed under
the
$SO(2,2)$ gauge-transformation
$$
U=\exp -t (J_{01}-J_{23}) \circ \exp r J_{03}$$
 to the $SO(2,2)$ flat connection \begin{eqnarray} A=
a (J_{02}+J_{13}) \end{eqnarray} for which the holonomy is
manifestly equal to
\begin{eqnarray} \exp \oint  [a (J_{02}+J_{13}) d\phi] = \exp 2 \pi a
(J_{02}+J_{13}) \end{eqnarray}

\section{Killing vectors.}

The self-dual metric has by construction four Killing vectors, namely
$\xi_{(1)}
\equiv \frac{\partial}{\partial \phi}$ belonging to the self-dual
$sl(2,\Re)$, and
the tree vectors $\eta_{(0)}, \eta_{(1)}, \eta_{(2)}$ belonging to the
anti-self-dual
$sl(2,\Re)$ and commuting therefore with $\xi_{(2)}$. In the
coordinates $(t, \phi,
r )$, these vectors are given explicitly by
\begin{eqnarray}
&&2\xi_{(1)}=
\frac{\partial}{\partial \phi} \label{5.1} \\
&& 2\eta_{(0)}= \frac{\partial}{\partial
t}  \label{5.2} \\
&& \eta_{(1)}=\frac{1}{2} \tanh 2r \cos 2t \frac{\partial}{\partial t}
+{ \cos 2t
\over 2 \cosh 2r} \frac{\partial}{\partial \phi} + \frac{1}{2} \sin 2t
\frac{\partial}{\partial r }         \label{5.3} \\
&&\eta_{(2)}= -\frac{1}{2}
\tanh 2r \sin 2t \frac{\partial}{\partial t} - { \sin 2t \over 2 \cosh
2r}
\frac{\partial}{\partial \phi} + \frac{1}{2} \cos 2t
\frac{\partial}{\partial r}
\label{5.4} \end{eqnarray}

One may ask whether there are any other independent Killing
vectors besides
(\ref{5.1})-(\ref{5.2}). It is easy to see that the answer to this
question is
negative.
Indeed, as shown  in \cite{BHTZ}, the problem amounts to
determining  all $SO(2,2)$ Killing
vectors $\eta$ of anti-de Sitter space that commute with the
$SO(2,2)$ matrix $\exp
2\pi \xi_{(1)}$, \begin{eqnarray} \label{5.5} \bigl[ \exp 2 \pi
\xi_{(1)}, \eta
\bigr] = 0 \end{eqnarray}

Now, the solutions of (\ref{5.5}) form a subalgebra of $so(2,2)$,
which contains $\Re
\oplus sl(2,\Re)$, where $\Re$ is the subalgebra of the ``self-dual''
$so(2,1)$
generated by $\xi_{(1)}$. There are only two subalgebras containing
$\Re \oplus
sl(2,\Re)$, namely, $\Re \oplus sl(2,\Re)$ itself or the full $so(2,2)$.
Since this
latter case is excluded (there exist elements of $so(2,2)$ that are
not invariant
by $\exp 2 \pi \xi_{(1)}$), we conclude that the isometry algebra of
the quotient
space is $\Re \oplus sl(2,\Re)$. There are thus only four independent
Killing vectors.

\section{Killing spinors.}

The metric (\ref{3.1}) can be viewed not just as a solution of Einstein
equation, but
also as a solution of adS supergravity with zero gravitini. As such, it
possesses
exact supersymmetries.

Exact supersymmetries are by definition  supersymmetry
transformations leaving the metric (\ref{3.1}) (with zero gravitini)
invariant. The
spinor parameters of these transformations solve the "Killing spinor
equation"
\begin{eqnarray} D_\lambda \psi = \frac{\epsilon}{2l}
\gamma_\lambda \psi,
\end{eqnarray} where $\epsilon = 1$ or $-1$ depending on the
representation of the
$\gamma$ matrices.

As is well known, there are two inequivalent two-dimensional
irreductible
representations of the $\gamma$ matrices in three spacetime
dimensions. One may be
taken to be $\gamma{(0)}=i\sigma^2, \gamma^{(1)}= \sigma^1$ and
$\gamma^{(2)}=\sigma^3$, where the $\sigma^k$ are the Pauli
matrices. The other is
given by $\gamma'^{(\lambda)}=-\gamma^{(\lambda)}$. We shall
consider here the
simplest supergravity model with negative cosmological constant
involving both
representations, namely, $(1,1)$ adS supergravity \cite{13}.

The anti-de Sitter metric $ds_{ads}^2= -dt^2+d\phi^2 + 2 \sinh 2r dt
d\phi + dr^2
 \ \ (-\infty<\phi<+\infty)$ possesses four Killing spinors, two for
each
inequivalent representation of the $\gamma$ matrices. In the tetrad
frame
\begin{eqnarray} h_{(0)}= \cosh r dt - \sinh r d\phi \nonumber  \\
h_{(1)}=
\sinh r dt + \cosh r d\phi \\ h_{(2)}= dr
 \nonumber  \end{eqnarray} the Killing spinors are given by
\begin{eqnarray}
\label{met} \psi = & \bigl[ [ (\epsilon-1) (A \cosh \phi + B \sinh
\phi)
e^{-r/2}+&(\epsilon+1)
 \sin t+K e^{r/2}] \nonumber \\
 & &\times (1+\gamma^{(2)}) \nonumber \\ & + [ (\epsilon-1)(A
\sinh \phi +B \cosh
\phi) e^{r/2}+ &(\epsilon+1)\cos t+K e^{-r/2}] \nonumber \\ &
&\times
(1-\gamma^{(2)}) \bigr] E
 \end{eqnarray} where $E$ is a constant spinor and $A$, $B$, $K$ are
constant.

Since the self-dual metric can be obtained  from $ds_{adS}^2$ by
making appropriate
identifications, it possesses locally as many Killing spinors as anti-de
Sitter
space. However, only a subset ot these Killing spinors are, in general,
compatible
with the identifications, i.e., invariant under the transformations of
the discrete
group used in the identifications. So, whereas all the local
integrability conditions
for the Killing equations are fulfilled \cite{CH}, there may be no
Killing spinor at
all because of global reasons.

Since the self-dual metric is obtained by making the identifications
$\phi \sim \phi
+ a \pi$, the only Killing spinors of the self-dual metric are those that
are
periodic or anti-periodic in $\phi$. By inspection of  (\ref{met}), one
finds that
these are given by: \begin{eqnarray} \bigl[ \sin (t+K) e^{r/2}
(1+\gamma^{(2)}) +
\cos (t+K) e^{-r/2} (1-\gamma^{(2)}) \bigr] E \end{eqnarray}

Thus, whereas anti-de Sitter space has four Killing spinors, the self-
dual geometry
has only two Killing spinors (two for one representation of the
$\gamma$ matrices and
zero for the other).

\section{String duality.}

The self-dual geometry (\ref{3.1}) may also be viewed as a solution
of the low energy
string equations in 2+1 dimensions \cite{Horowitz} \begin{eqnarray}
\label{eqnst}
\begin{array}{l}
S=\int {(-g)}^{\frac{1}{2}} e^{-\phi} \bigl[ \frac{4}{k} + R + 4 (\nabla
\phi)^2 -
\frac{1}{12} H_{\mu \nu \rho} H^{\mu \nu \rho} \bigr]
\\ R_{\mu \nu}+ 2
\nabla_\nu \nabla_\mu \phi - \frac{1}{4} H_{\mu \lambda \sigma}
{H_{\nu}}^{\lambda
\sigma}=0 \nonumber \\ \nabla^\mu ( e^{-\phi} H_{\mu \nu
\rho} ) =0  \\ 4
\nabla^2 \phi - 4 (\nabla \phi)^2 + \frac{4}{k} + R - \frac{1}{12} H^2
=0
\end{array}
\end{eqnarray} with zero dilaton $\phi=0$ and antisymmetric
tensor.
$B_{\lambda \mu}$ given by  \begin{eqnarray} B_{\phi t}
=\frac{a}{2} \sinh 2 r, \ \ \
B_{\phi r}= B_{t r} =0  \end{eqnarray} (so
 that $H_{\mu \nu \rho}= \frac{2}{l} \epsilon_{\mu \nu \rho} {(-
g)}^{\frac{1}{2}}
\not = 0$). In (\ref{eqnst}),  $k$ must be taken equal to $l^2$. Thus,
part of the
cosmological constant arises from the antisymmetric tensors and part
of it arises
from  the term $\frac{4}{k}$.

Given a solution of the low energy string equations with a Killing
vector, one may
construct by duality another solution of the same equations ( see
references therein
\cite{Horowitz,l}).
The duality transformation reads $(g_{\mu \nu}, B_{\mu  \nu },
\phi) \longrightarrow
({g '}_{\mu \nu}, {B'}_{\mu \nu},  \phi ')$  with  \begin{eqnarray}
 g_{xx} ' & =& 1/g_{xx}\ , \qquad  g_{xr} ' = B_{xr}/ g_{xx}\ ,
\nonumber\\
  g_{\alpha\beta}' & =& g_{\alpha\beta} - (g_{x\alpha}g_{x\beta} -
B_{x\alpha}B_{x\beta})/g_{xx}\ , \nonumber\\
  B_{x\alpha} ' & =& g_{x\alpha}/g_{xx}\ , \qquad
          B_{\alpha\beta} ' = B_{\alpha\beta} +2 g_{x[\alpha}
B_{\beta]x}/g_{xx}\ ,
\nonumber\\
 \phi  '& =& \phi - {1\over 2} \log |g_{xx}| \ .
 \label{transf} \end{eqnarray} where we assume that the solutions
$(g_{\mu \nu},
B_{\mu  \nu }, \phi)$ does not depend on $x$ and where $\alpha$
and $\beta$ runs over
all coordinates but $x$.

In \cite{Horowitz}, the transformation (\ref{transf}) was applied to
the $\phi$
translational symmetry of the black hole solutions to generate "black
string
solutions".
The Killing vector $\frac{\partial}{\partial \phi}$ is singled out  by
the fact that
it has closed orbits. In the cas of the black hole,
$\frac{\partial}{\partial \phi}$
is not self-dual and therefore, does not have a constant norm,
$g_{\phi\phi} \not =$
constant. As  a result, the transformation (\ref{transf}) generates a
different
geometry.

For the self-dual geometries, the Killing vector
$\frac{\partial}{\partial \phi}$
with closed orbits has the quite interesting property of being self-dual
and is thus of
constant norm. Accordingly, the transformation (\ref{transf}):
\begin{eqnarray} \label{transf1}
g'_{\phi\phi}=\frac{4}{a^2} & g'_{\phi t}= \frac{2}{a} \sinh 2r &
g'_{\phi r}=0
\nonumber \\ g'_{tt}=-1 & g'_{tr}=0 & g'_{rr}=0 \nonumber \\
B'_{t\phi}=\frac{1}{a}
\sinh 2r & B'_{tr}=0 & B'_{\phi r}=0 \nonumber \\ \Phi' = \Phi -
\frac{1}{2}
\log |a| \end{eqnarray} and simply amounts to replacing $a$ by
$\frac{4}{a}$. It does not
modify the geometry.

We can thus conclude that the self-dual geometries have also
remarkable duality
properties from the string duality point of view.

\section{Conclusions.}

We have analysed in this paper the self-dual geometries determined
from (2+1)-anti
de Sitter by identifications generated by a self-dual Killing vector.
We have shown that these geometries have quite interesting symmetry
properties: they have four Killing vectors
and two Killing spinors. They are also invariant under the string
duality transformation applied
to the angular translational symmetry $\phi \rightarrow \phi +
\alpha$.
\vskip 0.5 cm
\noindent {\bf Acknowledgments}
\vskip 0.3 cm
One of us (M.H.) is grateful
to Claudio Teitelboim and Jorge Zanelli for
their kind
invitation to the meeting ``The black hole 25 years after". This work
has been
supported  in part by a F.N.R.S. grant and by research contracts with
the
Commission of the European Community. O.C. is a
  ``chercheur I.R.S.I.A. ''.

\end{document}